\documentclass[12pt]{book}

\usepackage[dvips]{graphicx,color}
\usepackage{makeidx,tsukuba}

\makeauthorindex
\makeindex

\newcommand{\gsim}{\lower.7ex\hbox{$\;\stackrel{\textstyle>}{\sim}\;$}}
\newcommand{\lsim}{\lower.7ex\hbox{$\;\stackrel{\textstyle<}{\sim}\;$}}
\newcommand{\nuN}{$\nu N$}

\newcommand{\crs}{cross-section} 
\newcommand{\crss}{cross-sections} 

\newcommand{\stfs}{structure functions}

\begin{document}

\BookTitle{\itshape The 28th International Cosmic Ray Conference}
\CopyRight{\copyright 2003 by Universal Academy Press, Inc.}
%\tableofcontents
\pagenumbering{arabic}

\chapter{%   %%%%%%%%% <===== TITLE of the contribution
%%%%%%%%%%% The first letter of each word should be capital letter.
High Energy Neutrino Generator for Neutrino Telescopes
}

\author{%
Marek Kowalski and Askhat Gazizov \\
\emph{DESY Zeuthen, Platanenallee 6, D-15738 Zeuthen, Germany\\
}%% end of author
}

\section*{Abstract}
We present the high energy neutrino Monte Carlo event generator ANIS
(All Neutrino Interaction Simulation). The aim of the program is to
provide a detailed and flexible neutrino event simulation for high
energy neutrino detectors, such as AMANDA and ICECUBE. It generates
neutrinos of any flavor according to a specific flux, propagates them
through the Earth and in a final step simulates neutrino interactions
within a specified volume. All relevant standard model processes are
implemented. We discuss strength and limitations of the program, and
provide as an example event rates for atmospheric and $E^{-2}$
neutrino spectra. 

\section{Introduction}
The era of dedicated high-energy neutrino telescopes has just
begun. The currently operating detectors AMANDA and BAIKAL will hopefully 
soon be joined by  ANTARES, ICECUBE, NEMO and NESTOR. The unifying
goal is to detect neutrinos of extraterrestrial origin. For the Monte
Carlo (MC) simulation of atmospheric muons in such detectors, CORSIKA
is frequently used. The subsequent muon propagation may be done using
programs such as MMC, MUM or MUSIC. However, no standard for neutrino
generators has yet emerged. 

In this paper, the MC $\nu$-event generator \textbf{ANIS}
(\textbf{A}ll \textbf{N}eutrino \textbf{I}nteraction
\textbf{S}imulation) is described. It generates $\nu$-events of all
flavors, propagates them through the earth and finally simulates 
$\nu$-interactions within a specified volume around the detector. All
relevant SM processes, i.e.\ CC and NC $\nu N$-interactions as well as
resonant $\bar \nu_e e^-$-scattering are implemented. Neutrino
regeneration as expected in NC-scattering, $\nu + N \rightarrow \nu +
X$, and in $\tau$ production and decay chains, $\nu_\tau + N
\rightarrow \tau + X$,  $\tau \rightarrow \nu_\tau + (\nu_i) + X$, are
included in ANIS. 

In the next sections a  description of the characteristic
features of ANIS is presented. As an application of ANIS, the event
rates for atmospheric and astrophysical electron neutrinos (following
an $E^{-2}$ spectrum) are calculated.  

\section{The Signal Generator {\rm ANIS}}
ANIS is written in C++ and makes use of the CLHEP package [1]. The
internal event record is derived from the CLHEP \emph{HepMC} event
class. There are currently two output formats implemented: i) the
AMANDA specific event format \emph{f2000} [2] and ii) the HepMC
\emph{ascii} format [1]. 

The program is controlled through a steering file which allows to
enable specific interaction processes included in the
simulation. The currently implemented interaction channels include CC,
NC as well as resonant $\bar \nu_e e^-$-scattering. The cross-section
data for CC and NC reactions are provided through pre-calculated
external tables. The total cross-section is obtained through
interpolation. Large sets of possible final states, characterized by
the Bjorken variable $x$ and $y=E_h/E_\nu$, have been generated and
are randomly sampled from during generation of neutrino events.  The
use of pre-calculated tables makes the program fast and independent
of other packages. The cross-section data has been calculated up to
$10^{12}$~GeV and is discussed in more detail in the following
section.  

Primary $\nu$'s are randomly generated on the surface of the earth
with energy spectrum $F_\nu (E) \propto E^{-1}$ and are then propagated to 
the detector. Interacting with matter they are either absorbed (CC case) 
or regenerated at lower energies (NC case) [3]. In the special case of CC
$\nu_\tau N$-interaction a short-living $\tau$-lepton is produced. It
propagates in matter, loosing part of its energy, and finally 
decays giving rise to secondary $\nu_\tau$ and, in $\sim 17~\% $ of
the cases, to secondary $\nu_\mu$ or $\nu_e$. The $\tau$-decay in
ANIS has been simulated using TAUOLA [4]. Thereby the practically
full polarization of high energy $\tau$'s was taken into
account. Again, the previously generated decays of polarized
$\tau$-leptons are stored in tables, which are then sampled from by
ANIS. 
%Energy loss of the intermediate $\tau$'s is currently neglected,
%restricting the range of energies for accurate simulation of
%$\nu_\tau$ to about $10^{8}$ GeV. 

The density profile of the earth in ANIS is chosen according to the 
Preliminary Earth Model [5]. Regeneration effects are naturally
accounted for by feeding the secondary $\nu$'s back into the event
record. These $\nu$'s are assumed to be emitted parallel to the  
direction of the primary $\nu$. This is justified, since regeneration
effects become significant just at very high energies, where the
accumulated deflection angle is generally smaller than the telescope
angular resolution. 
  
Once the detection volume is reached, a final vertex is sampled 
along the $\nu$ trajectory  within the detection
volume (specified through the steering file). In the case of a CC
$\nu_\mu N$-interaction, ANIS correctly simulates the muon scattering
angle. Along with the full event, three weights for later use are
written out: the normalization of the flux, a weight for the
atmospheric flux [6], as well as a weight proportional to the total
cross-section of the neutrino interaction.

Event rates for atmospheric and various extraterrestrial neutrino
spectra  are obtained by applying the appropriate weights to the
events. This last  step is to be done by a user defined energy
dependent weight function during  analysis of the events in PAW, ROOT
or any other analysis programs. 

\section{Neutrino Cross Sections}
At $E_\nu \lsim 10$ PeV, deep inelastic $\nu N$-\crss\ may be
successfully described in the framework of pQCD. Parameterization of
the $\nu N$\ \stfs, $F_i^{\nu N}(x,Q^2)$, may be taken e.g.\ according to
CTEQ5 [7]. At higher energies these cross-sections are dominated by
scattering off small $x$ quarks. The unknown behavior of the \stfs\ at
$x \lsim 10^{-6}$ makes the calculations  model dependent. The
uncertainty in extrapolations of $F_i^{\nu N}(x,Q^2)$ to small $x$ and
large $Q^2$ influence the expected cross-sections at high energies. 

There are two possibilities presently included in ANIS: i) the smooth
both in $x$ and in $Q^2$ \emph{power-law} extrapolation of the pQCD CTEQ5
parameterization to small $x$ and large $Q^2$, and ii) \emph{hard
pomeron} [8] enhanced extrapolation [9]. The \crss\ of the first case,
denoted in the Fig.~1a as pQCD, practically coincide with [10], while
the second model (HP, dash-dotted curves) predicts approximately $2$
times higher \crss\ at $E_\nu \approx 1$ ZeV. The $y$-distributions,
normalized to the corresponding \crss, are plotted in Fig.~1b; the
solid and dashed curves stand for CC and NC $\nu N$-scattering,
respectively.      
     
\begin{figure}[t]
\begin{minipage}[b]{.5\linewidth}
  \centering
    \includegraphics[height=15pc]{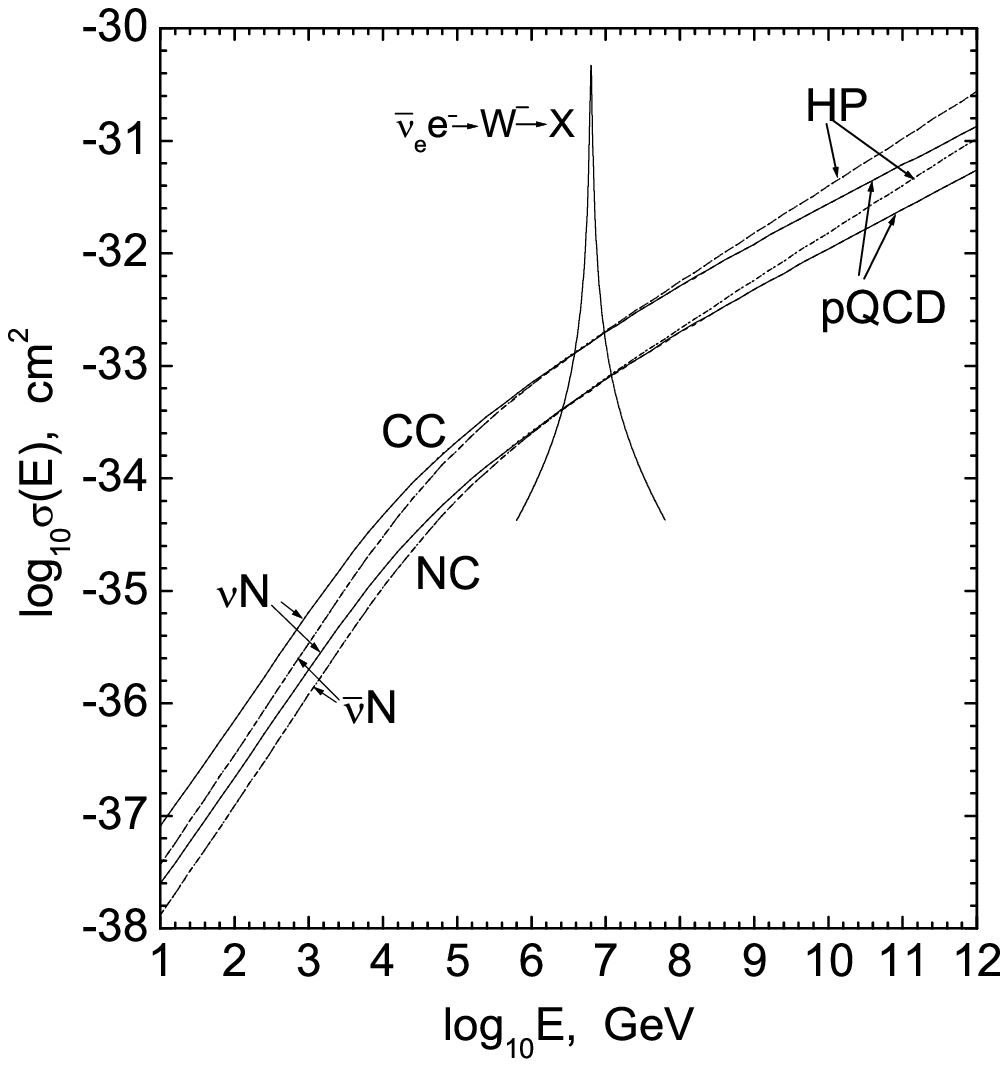}%
\end{minipage}\hfill
\begin{minipage}[b]{.5\linewidth}
  \centering
    \includegraphics[height=15pc]{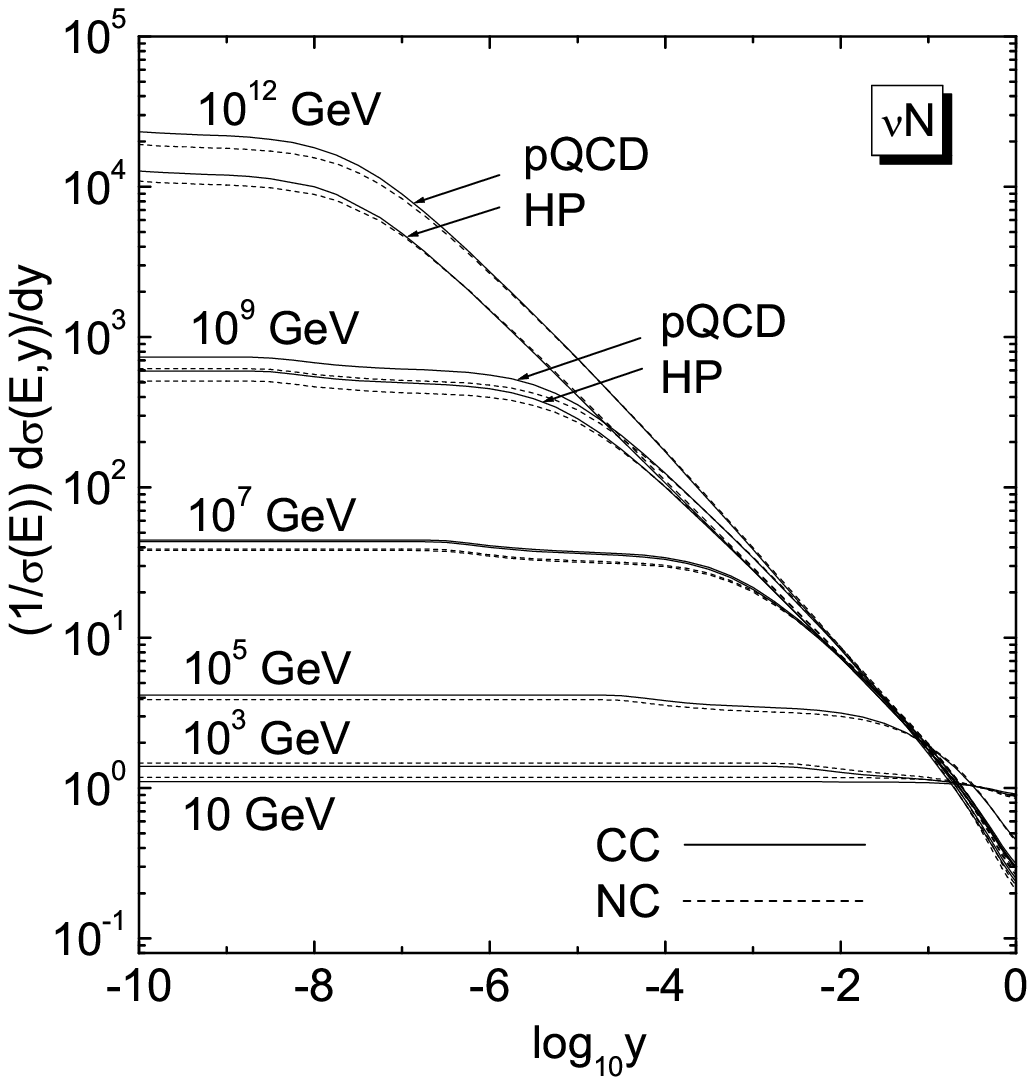}%
\end{minipage}
  
  \vspace{-0.5pc}
  \caption{a) \nuN- and resonance $\bar \nu_e e^- $-\crss\  and b)
  normalized $y$-distributions.} 
\end{figure}

The user can choose the desired  model through the steering file. It
is possible to add any new model to ANIS by providing the corresponding
\crs\ and final state tables.

Due to $m_e \ll m_N$,  at high-energies practically all $\nu
e^-$-\crss\ are negligible, with the only exception of $\bar \nu_e e^-
\rightarrow W^- \rightarrow anything$ at $E_{\bar \nu_e} \approx
6.3$~PeV [11] (see Fig.~1a). This resonance process is also included
in ANIS. Finally it should be noted, that ANIS can easily be extended
to include new processes beyond the Standard Model.
     
\begin{figure}[t]
\begin{minipage}[b]{.475\linewidth}
  \centering
    \includegraphics[height=14pc]{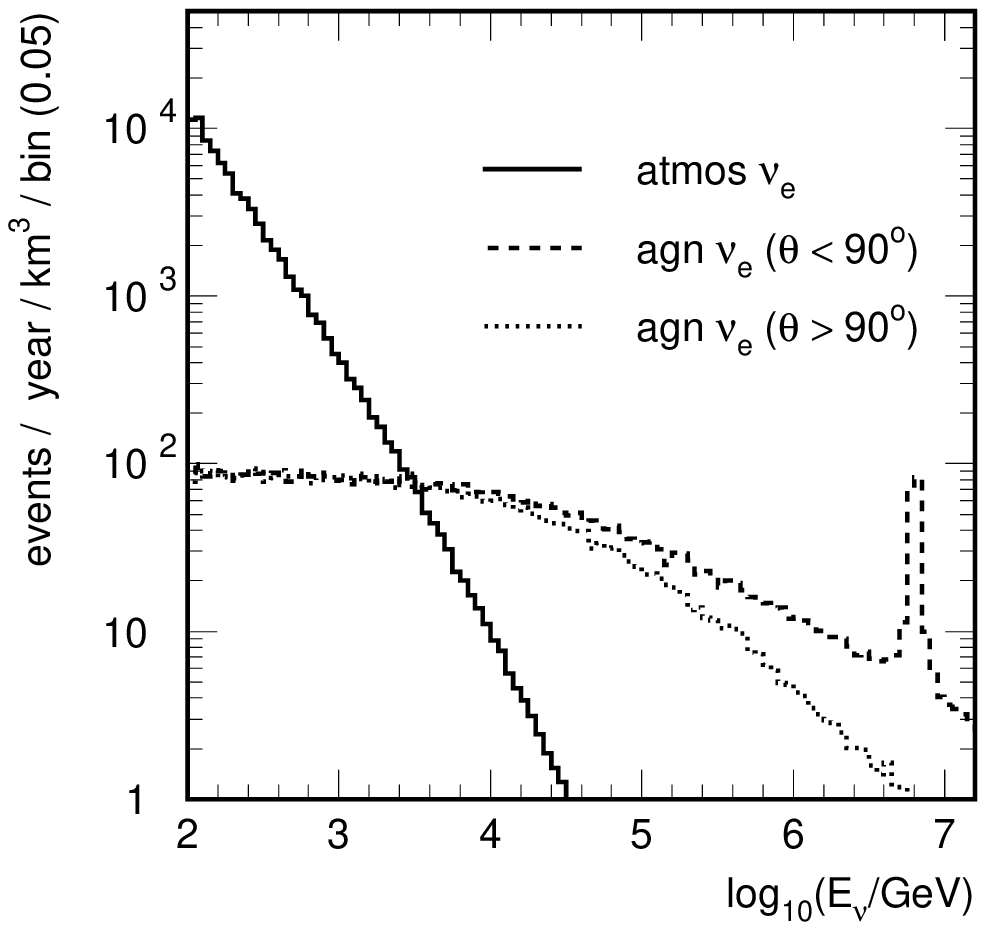}%
\end{minipage}
\begin{minipage}[b]{.475\linewidth}
  \centering
    \includegraphics[height=14pc]{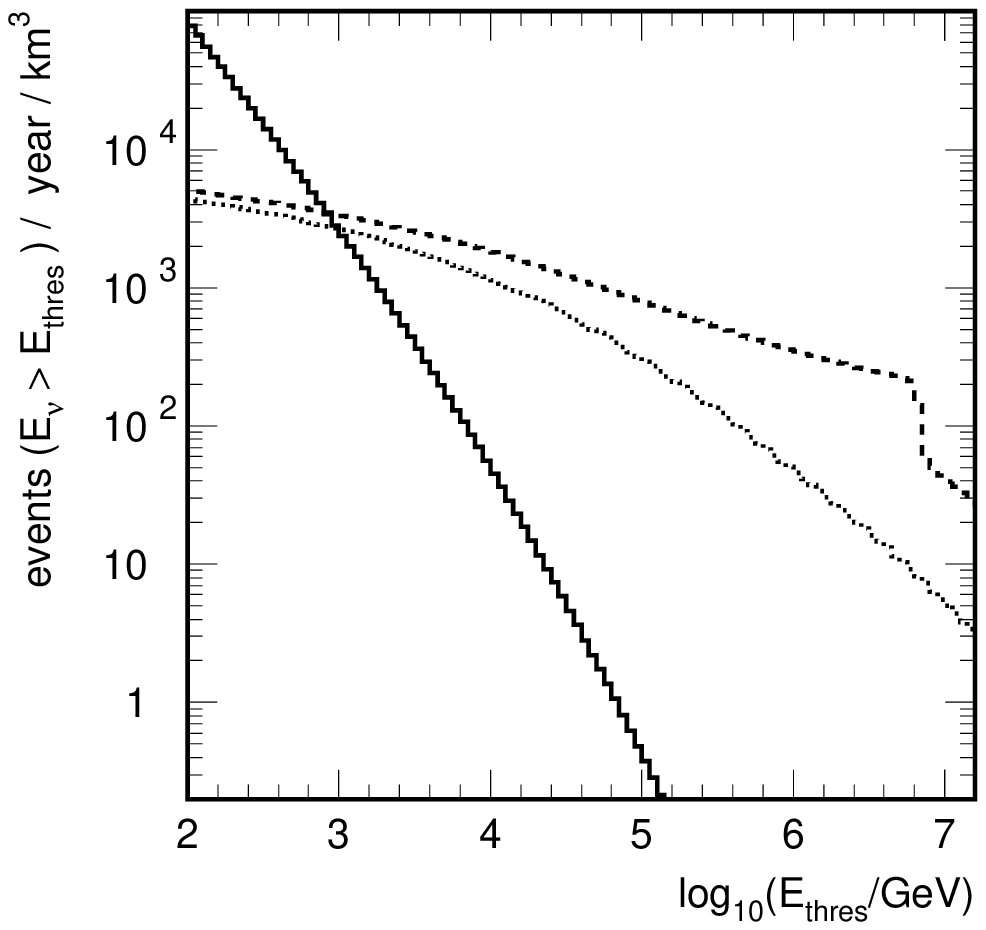}%  
\end{minipage}
  \vspace{-0.5pc}
  \caption{Number of $\nu_e +\bar \nu_e$ events per $({\rm km^3~year})$ are shown 
as a function of energy (left). The number of events above a threshold energy
${\rm E_{thres}}$ are shown in the right plot. }
\end{figure}

\section{Discussion}
As an application of ANIS,  
the resulting distributions of $10^5$ $\nu_e+\bar{\nu}_e$ events 
for an atmospheric spectrum and a diffuse AGN-like flux  
($F_\nu (E)= 1 \times 10^{-6} E^{-2}$ GeV$^{-1}$ cm$^{-2}$ s$^{-1}$
sr$^{-1}$)  are shown in Fig.~2. 
In order to illustrate the effects of $\nu$ propagation
through the earth, the contributions from the upper and lower
hemisphere are shown separately. Alternatively, $\nu_\mu$ and
$\nu_\tau$ can be simulated, however, muons produced in the reaction
have to be propagated with a separate program, e.g.\ with those
mentioned in the introduction. 
Concluding one can say, that the present version of ANIS allows the 
precise simulation of high-energy $\nu$-events of all flavors in a
wide energy range and designed
flexible to allow incorporation of possible new physics processes. 
\vspace{\baselineskip}
\re
~1.\ http://wwwinfo.cern.ch/asd/lhc++/clhep.
\re
~2.\ http://www-zeuthen.desy.de/nuastro/software/siegmund/f2000.ps.gz.
\re
~3.\ Berezinsky V. S. et al. 1986, Sov.\ J. Nucl.\ Phys.\ 43, 637.
\re
~4.\ Jadach S. et al.\ 1993, Comput.\ Phys.\ Commun.\ 76, 361.
\re
~5.\ see Ref.~[83] in Ghandhi R. et al. 1996, Astropart.\ Phys.\ 5, 81. 
\re
~6.\ Lipari P. 1993, Astropart.\ Phys.\ 1, 195.
\re
~7.\ CTEQ collab.: La H. L. et al. hep-ph/9903282;
http://www.phys.psu.edu/\~{}cteq/.
\re
~8. Donnachie A. and Landshoff P. V. 1998,  Phys.\ Lett.\ 437B, 408.
\re
~9.\ Gazizov A. Z., Yanush S. I. 2002, PRD 39, 941.
\re
10.\ Gandhi R. et al.\ 1998, Phys.\ Rev.\ D 58, 093009.
\re
11.\ Berezinsky V. S. and Gazizov A. Z. 1977, JETP Lett.\ 25, 276.
\endofpaper
\end{document}